\documentclass[3p,12pt]{elsarticle}

\usepackage[fleqn]{amsmath}
\usepackage{cases}
\usepackage{booktabs}
\usepackage{multirow}
\usepackage{xcolor}
\usepackage[normalem]{ulem}
\usepackage{tabularx}
\usepackage{siunitx}
\usepackage{comment}
\usepackage[ruled, linesnumbered]{algorithm2e}
\usepackage{algpseudocode}
\usepackage{algorithmicx}
\usepackage{grffile}
\usepackage{rotating}
\usepackage{lineno}
\usepackage{hyperref}

\usepackage{graphicx,enumerate}
\usepackage{amssymb}
\allowdisplaybreaks[4]
\usepackage{bm}
\usepackage{subfigure}
\usepackage{caption,color}
\usepackage{subeqnarray}
\usepackage{multirow}
\usepackage{lscape}
\usepackage{rotating}
\usepackage{graphics}
\usepackage{epstopdf}
\usepackage{threeparttable,booktabs,tabularx}
\usepackage{mathtools, empheq}

\journal{}

\usepackage{floatrow}
\usepackage{floatrow}\newfloatcommand{capbtabbox}{table}[][\FBwidth]
\floatname{algorithm}{Algorithm}
\biboptions{numbers,sort&compress} 

\begin{document}

\begin{frontmatter}
\title{
  Efficient solutions of eigenvalue problems in rarefied gas flows
}

 \author{Lei Wu\corref{mycorrespondingauthor}}
 \cortext[mycorrespondingauthor]{Corresponding authors:}
 \ead{wul@sustech.edu.cn}

\address{Department of Mechanics and Aerospace Engineering, Southern University of Science and Technology, Shenzhen 518055, China}

\begin{abstract}
The linear stability analysis of the Boltzmann kinetic equation has recently garnered research interest due to its potential applications in the high-altitude hypersonic flows, where rarefaction effects can render the Navier–Stokes equations invalid. Since the Boltzmann equation is defined in a seven-dimensional phase space, directly solving the associated eigenvalue problems is computationally intractable. 
In this paper, we propose an efficient iterative method to solve the linear stability equation of the kinetic equation. The solution process involves both outer and inner iterations. In the outer iteration, the shifted inverse power method is employed to compute selected eigenvalues and their corresponding eigenfunctions of interest. For the inner iteration, which involves inverting the high-dimensional system for the velocity distribution function, we adopt our recently developed general synthetic iterative scheme to ensure fast-converging and asymptotic-preserving properties. As a proof of concept, our method demonstrates both high efficiency and accuracy in planar sound wave and Couette flow. Each eigenpair can be computed with only a few hundred iterations of the kinetic equation, and the spatial cell size can be significantly larger than the molecular mean free path in near-continuum flow regimes. 
In particular, for the sound problem, we observe for the first time that in the transition regime (where the mean free path of gas molecules is comparable to the sound wavelength), large discrepancies arise among the results obtained from the Navier–Stokes equations, the Boltzmann equation with different viscosity indices (reflecting various intermolecular potentials such as hard-sphere, Maxwell, and shielded Coulomb interactions), and the simplified Shakhov kinetic model equation.

\end{abstract}


\end{frontmatter}
\section{Introduction}

There has been growing interest in rarefied gas flows, driven by advancements in modern engineering applications such as space exploration~\cite{ivanov1998computational,votta2013hypersonic}. These flows exhibit thermodynamic non-equilibrium behavior that cannot be accurately captured by the Navier–Stokes (NS) equations. Instead, the Boltzmann equation is employed to describe the gaseous system at the mesoscopic level. It models the streaming and collision of gas molecules using the velocity distribution function (VDF), which spans seven dimensions: time $t$, physical space $\bm{x}=(x_1,x_2,x_3)$, and molecular velocity space $\bm{v}=(v_1,v_2,v_3)$. According to the Chapman–Enskog expansion, the NS equations represent only a limiting form of the Boltzmann equation in the continuum flow regime~\cite{chapman1990mathematical}. 

Rarefied gas flows are characterized by the Knudsen number ($\text{Kn}$), defined as the ratio of the molecular mean free path to a characteristic flow length $L$. This dimensionless number is also proportional to the ratio of the Mach number to the Reynolds number. As a result, it has been widely assumed that flow instability does not arise in rarefied gas regimes, and studies have traditionally focused only on stationary solutions.
However, in hypersonic flight, strong compression leads to a flow regime ahead of the vehicle that is nearly continuum (small $\text{Kn}$), while the wake region behind the vehicle can approach the slip or even transition regime ($\text{Kn} \sim 1$). In such scenarios, flow instability and rarefaction effects may coexist in different regions and interact at their interfaces.

A linear stability analysis of the Boltzmann equation can be used to examine the stability of the base rarefied gas flow. Given the high-dimensional VDF, numerical simulation of the Boltzmann equation presents a significant challenge. Therefore, the stability of rarefied gas flows, e.g., the supersonic boundary-layer of a slightly rarefied gas, is analyzed based on the NS equation, with the velocity-slip and temperature jump conditions~\cite{OuCHen2021PoF,Levin2022}. This pose little technique challenges as the linear stability equation (LSE) of NS equations are well studied~\cite{Malik1990JCP,Zhong2024JCP,Dong2021AIAA}.

Recently, Zou \textit{et al} analyzed the stability of the Poiseuille and Couette flows~\cite{Zou2022PoF,Zou2023JFM} by directly solving the LSE of the simplified BGK kinetic model~\cite{BGK1954}. This approach is feasible because: (i) the three-dimensional velocity space is reduced to two dimensions by assuming homogeneity in the third direction; (ii) when the Knudsen number and flow velocity are not large, the molecular velocity space can be discretized using Gauss–Hermite quadrature with relatively few nodes; and (iii) the LSE is formulated in only one spatial dimension. As a result, the total number of discretized spatial and velocity grid points is approximately $100\times10\times10$, allowing all the eigenpairs to be computed within minutes. 
However, for hypersonic flows around the Apollo reentry capsule or X-38 space vehicles, simulations typically involve millions of spatial cells and tens of thousands discretized velocity points~\cite{zhang2024efficient}. This raises a critical question: how can we develop efficient numerical methods for solving the LSE of the kinetic equation under such demanding conditions? 

Note that a similar $k$-eigenvalue problem has been extensively studied in the context of the neutron transport equation~\cite{Alcouffe01101977,ADAMS20023}, where the positive largest eigenvalue is used to determine whether the chain reaction is subcritical, critical, or supercritical—corresponding to decay, steady-state operation, and unsafe power increase, respectively. This problem is typically solved using the power iteration method, rather than by directly solving the extremely large linear systems.

Theoretically, the power method can be employed to compute the eigenpairs of the LSE of the Boltzmann equation. However, significant challenges arise in near-continuum regimes with small Knudsen numbers, where frequent molecular collisions impede the convergence of iterations—making both the base flow and the eigenpairs difficult to obtain. For instance, in the one-dimensional Poiseuille flow with $\text{Kn}\sim10^{-3}$, the conventional iterative scheme (CIS) requires approximately one million steps to converge. Worse still, the resulting solution is highly susceptible to numerical dissipation if the spatial grid is not sufficiently refined~\cite{wang2018comparative}. 
To the best of our knowledge, the power method has not been previously applied to find the eigenpairs of the Boltzmann equation in rarefied gas dynamics.

Recently, we have developed the general synthetic iterative scheme (GSIS) to efficiently and accurately obtain base flow solutions of the Boltzmann equation—within only dozens of iterations, even on coarse spatial grids~\cite{su2020can,su2020fast,liu2024further}. This method demonstrates several orders of magnitude improvement in computational time compared to the widely used Direct Simulation Monte Carlo method~\cite{zhang2024efficient}.
The fast convergence and asymptotic-preserving properties of GSIS are achieved through a macroscopic synthetic equation, which incorporates both the continuum-level constitutive relations (Newton’s law of viscosity and Fourier’s law of heat conduction) and higher-order corrections derived directly from the kinetic equation to account for rarefaction effects. These macroscopic equations are relatively easy to solve and serve to efficiently guide the VDF toward its steady-state solution.


In the present work, we shall develop a numerical method to solve the LSE of the kinetic equation, by combining the shifted inverse power method and GSIS. 
Rather than computing all eigenvalues, we focus on the least stable mode and its nearby modes, which are typically of primary interest in most applications.
The remainder of this paper is organized as follows. Section~\ref{sec:2} introduces the Boltzmann equation along with its simplified Shakhov and BGK kinetic model equations. Section~\ref{sec:LSE} present the derivation of LSE for the kinetic equations. As a proof of concept, section~\ref{sec:sound} details the numerical method for the LSE to explore the properties of planar sound waves, especially the role of intermolecular potential which is not included in the Boltzmann equation. Section~\ref{sec:Couette} presents the numerical computation of eigenpairs for Couette flow. Finally, conclusions and outlooks are summarized in section~\ref{sec:conclusion}.

\section{Kinetic equations}\label{sec:2}

In gas kinetic theory, the VDF $f(t, \bm{x}, \bm{v})$ is used to describe the state of monatomic gas. The macroscopic quantities are obtained by taking moments of VDFs, e.g., the mass density $\rho$, flow velocity $\bm{u}$, traceless stress ${\sigma}_{ij}$, temperature $T$, and heat flux $\bm{q}$, are
\begin{equation}\label{eq:getmoment}
    \left(\rho,~\rho\bm{u},~{\sigma}_{ij},~\frac{3}{2}\rho T,~\bm{q}\right)=\int \left(1,~\bm{v},~2c_ic_j-\frac{2}{3}c^2\delta_{ij}, ~c^2, ~c^2\bm{c}
        \right) f \mathrm{d}\bm{v},
\end{equation}
where $\bm{c}=\bm{v}-\bm{u}$ is the thermal (peculiar) velocity, $c$ is the thermal speed, $\delta_{ij}$ is the Kronecker delta function, the subscripts $i,j=1,2$, or 3 denotes the spatial directions, and the integration is performed over the entire velocity space. The local Maxwellian equilibrium distribution is expressed as 
\begin{equation}
    F_{eq}(\rho,\bm{u},T)=\frac{\rho(t,\bm{x})}{\left[\pi T(t,\bm{x})\right]^{3/2} } \exp\left[-\frac{|\bm{v}-\bm{u}(t,\bm{x})|^2}{T(t,\bm{x})}\right].
\end{equation}

Note that we have used dimensionless variables in the above equations. Specifically, the spatial variable $\bm{x}$ is normalized by the characteristic flow length $L$, and the temperature by the reference temperature $T_0$. The velocity is scaled by the most probable speed $v_m = \sqrt{2k_B T_0}/m$, where $k_B$ and $m$ denote the Boltzmann constant and molecular mass, respectively. Time is normalized by $L/v_m$, and density by the reference density $\rho_0$. The pressure and stress is normalized by $p_0 = \rho_0 k_B T_0 / m$, and the normalized static gas pressure is $\rho T$. The heat flux is normalized by $p_0 v_m$.

\subsection{The Boltzmann equation}
The Boltzmann equation reads \cite{Wu2013JCP, Wu2014JFM}
\begin{equation}\label{Boltzmann}
     \frac{\partial f}{\partial t}+ \bm{v}\cdot\frac{\partial f}{\partial \bm{x}}=Q(f,f),
\end{equation}
which describes the streaming (left-hand-side of the equation) and binary collision between monatomic gases (right-hand-side). The Boltzmann collision operator $Q(f,f)=Q^+-\nu {f}$ consists of the gain term
\begin{equation}\label{coll_gain_normalization}
	Q^+(f,f)=\iint B(\theta,|\bm{v}-\bm{v}_\ast|)
	{f}({\bm{v}}'_{\ast}){f}({\bm{v}}')d\Omega d{\bm{v}}_\ast,
\end{equation}
and the loss term $\nu f$, with $\nu$ being the collision frequency
\begin{equation}\label{coll_fre}
	\nu(f)=\iint B(\theta,|\bm{v}-\bm{v}_\ast|)
	{f}({\bm{v}}_{\ast})d\Omega d{\bm{v}}_\ast.
\end{equation}

In the Boltzmann collision operator, the subscript $\ast$ represents the second molecule in the binary collision, while the superscript prime stands for quantities after the collision; the post-collision velocities are related to the pre-collision velocities as
\begin{equation}
\begin{aligned}
	\bm{v}' &= \frac{\bm{v} + \bm{v}_*}{2} + \frac{|\bm{v} - \bm{v}_*|}{2} \boldsymbol{\Omega}, \\
	\bm{v}_*' &= \frac{\bm{v} + \bm{v}_*}{2} - \frac{|\bm{v} - \bm{v}_*|}{2} \boldsymbol{\Omega},
\end{aligned}
\end{equation}
where $\bm{v}-\bm{v}_\ast$ is the relative pre-collision velocity and $\boldsymbol{\Omega}$ is a vector in the unit sphere along the relative post-collision velocity $\bm{v}'-\bm{v}'_\ast$. The deflection angle $\theta$ between the pre- and post-collision relative velocities satisfies $\cos\theta=\boldsymbol{\Omega}\cdot(\bm{v}-\bm{v}_\ast)/|\bm{v}-\bm{v}_\ast|$.

The inverse power-law potential is considered this paper, in which the intermolecular force decays with distance according to a power law with exponent $\eta$. Consequently, the shear viscosity $\mu$ of the gas can be expressed as
\begin{equation}
    \mu(T)=\mu(T_0) \left( \frac{T}{T_0} \right)^\omega,
\end{equation}
with $\omega=(\eta+3)/2(\eta-1)$ being the viscosity index. For Maxwellian gas, we have $\eta=5$ and $\omega=1$, while for hard-sphere gas $\eta$ is infinity and $\omega=0.5$; the Coulomb potential has $\eta=2$ and $\omega=2.5$. However, due to the Debye shielding effect, the effective viscosity index is less than 2.5~\cite{Su2019JCP}.  The Boltzmann collision kernel is modeled as~\cite{Wu2013JCP}
\begin{equation}
 B(\theta,|\bm{v}-\bm{v}_\ast|)  =\frac{5}{64\sqrt{2}\Gamma^2\left(\frac{7+\alpha}{4}\right)
	\text{Kn}} |\bm{v}-\bm{v}_\ast|^\alpha\sin^{\frac{\alpha-1}{2}}{\theta},
\end{equation}
where $\alpha=2(1-\omega)$, and 
the Knudsen number  is 
\begin{equation}\label{Knudsen}
 \text{Kn}=
     \frac{\mu(T_0)}{p_0L}\sqrt{\frac{\pi k_B T_0}{2m}},
\end{equation}
where $\Gamma$ is the gamma function.


\subsection{The simplified kinetic model equations}

The Boltzmann collision operator is hard to solve deterministically. Therefore, in many engineering applications, the following kinetic model is used: 
\begin{equation}\label{Shakhov}
   \frac{\partial f}{\partial t}+ \bm{v}\cdot\frac{\partial f}{\partial \bm{x}}=\delta_{rp} 
   \rho{}T^{1-\omega}\left(L^+_sf-f\right),
\end{equation}
where the gain term (or the reference VDF to which the VDF relaxes) is
\begin{equation}\label{Ref_Shakhov}
    L^+_sf=F_{eq}(\rho,\bm{u},T)
    \left[1+\frac{4(1-\text{Pr})}{5}\frac{\bm{q}\cdot
	\bm{c}}{\rho{}T^2}\left(\frac{c^2}{T}-\frac{5}{2}\right)\right],
\end{equation}
and the rarefaction parameter is inversely proportional to the Knudsen number as
\begin{equation}
    \delta_{rp}=\frac{\sqrt\pi}{\text{2Kn}}.
\end{equation}

For a monatomic gas, the Prandtl number $\text{Pr}$ is very close to $2/3$, as confirmed by both experimental measurements and the Chapman–Enskog expansion of the Boltzmann equation~\cite{chapman1990mathematical}. In the Shakhov model, we have $\text{Pr} = 2/3$. When $\text{Pr} = 1$, the Shakhov model reduces to the BGK model. While this simplified model significantly facilitates numerical simulations, the full Boltzmann collision operator requires more advanced techniques, such as the fast spectral method~\cite{Wu2013JCP}.

\subsection{The kinetic boundary condition}

The kinetic boundary condition describes how gas molecules are reflected upon colliding with a solid wall. Here, we consider the Maxwell diffuse boundary condition, in which the incident gas molecules are re-emitted into the fluid domain in thermal equilibrium with the wall's temperature $T_w$ and velocity $\bm{u}_w$. Assuming that the unit normal vector to the solid wall, pointing into the fluid domain, is $\bm{n}_w$, the reflected VDF satisfies
\begin{equation}\label{Maxwell_boundary}
f_\text{reflected} = \frac{\rho_{w}}{(\pi T_w )^{3/2}} \exp\left(-\frac{|\bm{v}-\bm{u}_w|^2}{T_w} \right), \quad \text{for } \bm{n}_w\cdot(\bm{v}-\bm{u}_w) > 0, 
\end{equation}
where $\rho_{w} = -2\sqrt{\pi/T_w} \int_{\bm{n}_w\cdot(\bm{v}-\bm{u}_w)< 0}  \bm{n}_w\cdot(\bm{v}-\bm{u}_w) f_\text{incident} d\bm{v}$ is determined by the non-penetration condition.

\section{The linear stability equation}\label{sec:LSE}

In this section, we first introduce the base flow and its numerical method. We then derive the LSE by linearizing around the base VDF and the macroscopic quantities.

\subsection{Base flow}

The base flow is a stationary solution, where the VDF,  denoted by $f_b(\bm{x},\bm{v})$, satisfying the following kinetic equations
\begin{equation}\label{Shakhov_base}
   \bm{v}\cdot\frac{\partial f_b}{\partial \bm{x}}=
   \delta_{rp}
   \rho_b{}T_b^{1-\omega}
   (L^+_{s}f_b-f_b),
\end{equation}
and
\begin{equation}\label{Boltzmann_base}
   \bm{v}\cdot\frac{\partial f_b}{\partial \bm{x}}=
   Q^+(f_b,f_b)-\nu(f_b) f_b,
\end{equation}
for the Shakhov and Boltzmann equations, respectively,
where the macroscopic quantities of the base flow are 
\begin{equation}\label{moment_base}
    \left(\rho_b,~\rho_b\bm{u}_b,~{\sigma}_{b,ij},~\frac{3}{2}\rho_b T_b,~\bm{q}_b\right)=\int\left(1,~\bm{v},~2c_{b,i}c_{b,j}-\frac{2}{3}c_b^2\delta_{ij}, {c_b^2}, c_b^2\bm{c}_b
        \right) f_b \mathrm{d}\bm{v},
\end{equation}
and $\bm{c}_b=\bm{v}-\bm{u}_b$ is the thermal (peculiar) velocity relative to the base flow velocity $\bm{u}_b$.
The boundary condition is described by Eq.~\eqref{Maxwell_boundary} when $f$ is replaced by $f_b$. 

In numerical simulations, the spatial variable $\bm{x}$ and the molecular velocity space $\bm{v}$ are discretized. The streaming term can be treated using finite-difference, finite-volume, or discontinuous Galerkin methods, while the velocity-space integration is handled using Gauss–Hermite quadrature or Newton–Cotes quadrature. In principle, the base flow can be obtained by transforming the discretized forms of Eqs.~\eqref{Shakhov_base} and~\eqref{Boltzmann_base} into extremely large linear systems for $f_b$. However, this approach is computationally prohibitive~\cite{Luc2020JCP}. As a result, these differential-integral kinetic equations are typically solved using the CIS~\cite{Sone1989PoFA,Wu2013JCP}, given by (take the Boltzmann equation for an example):
\begin{equation}\label{Boltzmann_base_CIS}
   \nu\left(f^{(k)}_b\right) f^{(k+1)}_b+\bm{v}\cdot\frac{\partial f^{(k+1)}_b}{\partial \bm{x}}=
   Q^+\left(f^{(k)}_b,f^{(k)}_b\right),
\end{equation}
where $k$ is the iteration number. Starting from an initial guess of the VDF, the iteration continues until the relative difference in macroscopic quantities falls below a specified criterion. Normally, the CIS is efficient when the Knudsen number is large, as the steady-state solution can be obtained within dozens of iterations. However, in the near-continuum regime, the required number of iterations becomes extremely large, e.g., up to one million when $\text{Kn} \sim 10^{-3}$~\cite{wang2018comparative}. To address this, the GSIS was recently proposed, reducing the iteration count to just a few dozen across all flow regimes~\cite{zhang2024efficient,su2020can}. We will not go into detail here, as the following two sections will be devoted to developing the GSIS for the LSE.



\subsection{Linearization around the base flow}

To perform the linear stability analysis, the VDF is expressed as 
\begin{equation}\label{vdf_perturbation}
    f(t, \bm{x}, \bm{v})=f_b(\bm{x}, \bm{v})+\tilde{f}(t, \bm{x}, \bm{v}),
\end{equation}
where $\tilde{f}(t, \bm{x}, \bm{v})$ is the perturbed VDF satisfying $|\tilde{f}(t, \bm{x}, \bm{v})/f_b(\bm{x}, \bm{v})|\ll1$. Accordingly, the macroscopic quantities $M$ can be expressed as $M=M_b+\tilde{M}$, where $|\tilde{M}/M_b|\ll1$. Specifically, the perturbed density, velocity, temperature, stress, and heat flux are given by  
\begin{equation}\label{perturbation_quantities}
\begin{aligned}
    \tilde{\rho}(t, \bm{x})=&\int \tilde{f}(t, \bm{x}, \bm{v}) d\bm{v}, \\
    \tilde{\bm{u}}(t, \bm{x})=&\frac{1}{\rho_b(\bm{x})} \int \bm{c}_b \tilde{f}(t, \bm{x}, \bm{v}) d\bm{v}, \\
    \tilde{T}(t, \bm{x})=&  \frac{2T_b(\bm{x})}{3\rho_b(\bm{x})}\int{}\left(\frac{{c}_b^2}{T_b(\bm{x})}-\frac{3}{2}\right)  
    \tilde{f}(t, \bm{x}, \bm{v}) d\bm{v},\\
    \tilde{\sigma}_{ij}(t, \bm{x})=& 2\int{} \left( c_{bi}c_{bj}-\frac{c_b^2}{3}\delta_{ij}\right)
    \tilde{f}(t, \bm{x}, \bm{v}) d\bm{v},\\
    \tilde{\bm{q}}(t, \bm{x})
    =&T_b(\bm{x})\int{} \left(\frac{c_b^2}{T_b(\bm{x})}-\frac{5}{2}\right)\bm{c}_b  \tilde{f}(t, \bm{x}, \bm{v}) d\bm{v}.
\end{aligned}
\end{equation}

Substituting Eq.~\eqref{vdf_perturbation} into Eq.~\eqref{Boltzmann}, we obtain the following evolution equation of the perturbed VDF in the linearized Boltzmann equation:
\begin{equation}\label{perturbation_Boltzmann}
\begin{aligned}
   \frac{\partial \tilde{f}}{\partial t}+ \bm{v}\cdot\frac{\partial \tilde{f}}{\partial \bm{x}}=
   \iint & B(\theta,|\bm{v}-\bm{v}_\ast|) \\
\times &\left[
      \tilde{f}({\bm{v}}'_{\ast}){f_b}({\bm{v}}')
    +{f}_b({\bm{v}}'_{\ast})\tilde{f}({\bm{v}}')
    -\tilde{f}({\bm{v}}_{\ast}){f_b}({\bm{v}})
    -{f}_b({\bm{v}}_{\ast})\tilde{f}({\bm{v}})
    \right]d\Omega d{\bm{v}}_\ast.
\end{aligned}
\end{equation}

Similarly, substituting Eq.~\eqref{vdf_perturbation} into Eq.~\eqref{Shakhov}, we obtain the following evolution equation of the perturbed VDF in the linearized Shakhov model:
\begin{equation}\label{perturbation_Shakhov}
\begin{aligned}
   \frac{\partial \tilde{f}}{\partial t}+ \bm{v}\cdot\frac{\partial \tilde{f}}{\partial \bm{x}}=&\delta_{rp} 
   \rho_bT_b^{1-\omega}(L^+_p\tilde{f}-\tilde{f})
   +\delta_{rp} 
   \rho_bT_b^{1-\omega}\left[\frac{\tilde{\rho}}{\rho_b}+(1-\omega)\frac{\tilde{T}}{T_b} \right]
   \left(L^+_{s}f_b-f_b\right),
\end{aligned}
\end{equation}
where
\begin{equation}\label{perturbation_Shakhov1}
\begin{aligned}
   L^+_p\tilde{f}=& F_{eq}(\rho_b,\bm{u}_b,T_b)\left[
   \frac{\tilde{\rho}}{\rho_b}+\frac{2}{T_b}\bm{c}_b\cdot\tilde{\bm{u}}
   +\frac{\tilde{T}}{T_b}\left(\frac{c_b^2}{T_b}-\frac{3}{2}\right)
   \right]\\
   +&F_{eq}(\rho_b,\bm{u}_b,T_b)\frac{4(1-\text{Pr})}{5\rho_bT_b^2}\left(\frac{c_b^2}{T_b}-\frac{5}{2}\right) 
   \left[
    \bm{c}_b\cdot\tilde{\bm{q}}-\bm{q}_b\cdot\tilde{\bm{u}} 
    -\bm{q}_b\cdot \bm{c}_b
    \left(\frac{\tilde{\rho}}{\rho_b}+2\frac{\tilde{T}}{T_b}\right)
    \right]\\
    -&F_{eq}(\rho_b,\bm{u}_b,T_b)\frac{4(1-\text{Pr})}{5{\rho_bT_b^2}}{\bm{q}_b\cdot
	\bm{c}_b}\left[
    \frac{2}{T_b}\bm{c}_b\cdot\tilde{\bm{u}}+\frac{c^2_b}{T^2_b} \tilde{T}
    \right].
\end{aligned}    
\end{equation}

Following Eq.~\eqref{Maxwell_boundary}, the kinetic boundary condition for the perturbed VDF is 
\begin{equation}
\begin{aligned}
    \tilde{f}_\text{reflected}=&{f}_\text{reflected}-{f}_\text{b,reflected}\\
    =&\frac{\tilde{\rho}_{w}}{(\pi T_w )^{3/2}} \exp\left(-\frac{|\bm{v}-\bm{u}_w|^2}{T_w} \right), \quad \text{for } \bm{n}_w\cdot(\bm{v}-\bm{u}_w)> 0, 
\end{aligned}
\end{equation}
where $\tilde{\rho}_{w} = -2\sqrt{\pi/T_w} \int_{\bm{n}_w\cdot(\bm{v}-\bm{u}_w) < 0} 
\tilde{f}_\text{incident}  d\bm{v}$.

\subsection{Eigenvalue problems}
One may analyze the temporal instability problem by assuming a planar sound wave perturbation like 
\begin{equation}
    \tilde{f}=\hat{f}(x_1,x_3,\bm{v}) 
    \exp(i{K}x_1-i\varpi{t}),
\end{equation}
where the real number ${K}$ is the wavenumber the planar sound wave, and the complex number $\varpi$ is the frequency (real part) and growth rate (imaginary part). Accordingly, the macroscopic quantities in Eq.~\eqref{perturbation_quantities} are expressed as $\tilde{M}=\hat{M} \exp(i{K}x_1-i\varpi{t})$. In this case, the streaming operator in Eqs.~\eqref{perturbation_Boltzmann} and \eqref{perturbation_Shakhov} is replaced by
\begin{equation}\label{plane_wave}
     \frac{\partial \tilde{f}}{\partial t}+ \bm{v}\cdot\frac{\partial \tilde{f}}{\partial \bm{x}}
     \rightarrow
     -i\varpi \hat{f}+i{K}v_1\hat{f}
     +v_2\frac{\partial \hat{f}}{\partial x_2}
     +v_3\frac{\partial \hat{f}}{\partial x_3},
\end{equation}
and the tildes in VDF and corresponding macroscopic quantities are replaced by hats. For simplicity, we decompose the collision operator into the gain term $L^+\hat{f}$ and loss term $\nu \hat{f}$. For example, in the linearized Boltzmann equation we have $\nu=\nu(f_b)$, while in the Shakhov model $\nu=\delta_{rp}$, and all others terms in the collision operator belong to the gain term.  Eventually, we have the eigenvalue problems in high-dimensional space:
\begin{equation}\label{general_eigen_problem}
-i\varpi \hat{f}=-\left(i{K}v_1\hat{f}
     +v_2\frac{\partial \hat{f}}{\partial x_2}
     +v_3\frac{\partial \hat{f}}{\partial x_3}\right)
     +\left( L^+\hat{f}-\nu \hat{f} \right).
\end{equation}

We note that this problem is similar to the $k$-eigenvalue formulation in neutron transport~\cite{Alcouffe01101977,ADAMS20023}. However, key differences exist between the $k$-eigenvalue problem and LSE of rarefied gas flows. First, in neutron transport, all eigenvalues are real, and only the one with the largest magnitude is of practical significance, while in rarefied gas flows the eigenvalues are generally complex. Second, the eigenvalue appears in the source term related to the neutron intensity, a macroscopic quantity. In contrast, in LSE of rarefied gas flows, the eigenvalue appears in a term multiplied by the VDF, specifically in the form $-i\varpi \hat{f}$.

To avoid computing eigenvalues by directly solving large linear systems, we employ the shifted inverse power method in combination with the GSIS to achieve efficient computation. Step-by-step details of the approach will be provided below. We first examine the simplest case—sound wave propagation—to illustrate the core concept, and then proceed to the LSE of the Couette flow. If the proposed method proves effective in these two cases, it can be extended to more complex scenarios involving intricate flow configurations.

\section{Planar sound wave}\label{sec:sound}

Consider the planar sound wave traveling in the infinite $x_2$ domain, which takes the form of $\exp(iKx_2-i\varpi{t})$.
The characteristic length $L$ is chosen to make the rarefaction parameter $\delta_{rp}=1$. We first consider the linearized Shakhov model for planar sound waves, which can be quickly implemented by anyone with basic knowledge of numerical analysis to test the performance of our method. 

Since the base flow is in global equilibrium with the macroscopic quantities 
\begin{equation}\label{sound_base_flow}
    \rho_b=\rho_0, ~T_b=T_0, ~\bm{u}_b=0, ~\bm{\sigma}_b=0, ~\bm{q}_b=0, 
\end{equation}
we have $\bm{c}_b=0$,
\begin{equation}
    f_b=f_{eq}=\frac{\exp(-v^2)}{\pi^{3/2}},
\end{equation}
the term $L^+_{s}f_b-f_b$ in Eq.~\eqref{perturbation_Shakhov} is zero, and the term 
$\left[
    \bm{c}_b\cdot\hat{\bm{q}}-\bm{q}_b\cdot\hat{\bm{u}} 
    -\bm{q}_b\cdot \bm{c}_b
    \left(\frac{\hat{\rho}}{\rho_b}+2\frac{\hat{T}}{T_b}\right)
    \right]$
in the middle line of Eq.~\eqref{perturbation_Shakhov1} becomes $\bm{v}\cdot\bm{\hat{q}}$.
Therefore, we have 
\begin{equation}\label{Shakhov_sound}
    -i\varpi{}\hat{f}+iKv_2\hat{f}=L^+_s\hat{f}-\hat{f},
\end{equation}
where the gain term is simplified to 
\begin{equation}
L^+_s\hat{f}=\left[\hat{\rho}+2\hat{u}_2v_2+\hat{T}\left(v^2-\frac{3}{2}\right)+\frac{4}{15}\hat{q}_2v_2\left(v^2-\frac{5}{2}\right)\right]f_{eq},
\end{equation}
with 
\begin{equation}\label{MP}
[\hat{\rho},~\hat{u}_2,~\hat{T},~\hat{q}_2]=
\int\left[1,~v_2,~\frac{2}{3}\left(v^2-\frac{3}{2}\right),~\left({v^2}-\frac{5}{2}\right)v_2\right]\hat{f}d\bm{v}.
\end{equation}

Given the perturbation wavenumber $K$, we aim to find the eigenvalue 
$-i\varpi$ in the kinetic equation~\eqref{Shakhov_sound}. To avoid working with large matrices in the linear algebra computations, we employ the shifted inverse power method to find $\hat{f}^{(m+1)}$ when $\hat{f}^{(m+1)}$ is known:
\begin{equation}\label{Shakhov_sound_si_power}
    (L^+_s-1-iKv_2-p)\hat{f}^{(m+1)}=\hat{f}^{(m)},
\end{equation}
where $m$ is the outer iteration number and $p$ is the estimated eigenvalue. When $\hat{f}^{(m+1)}$ is solved, the VDF is normalized after each outer iteration using the value with the maximum magnitude, denoted as $\lambda^{(m+1)}$. During iteration, this will form a consequence of eigenvalues: 
\begin{equation}\label{eig_power_itr}
    -i\varpi^{(m+1)}=p+\frac{1}{\lambda^{(m+1)}}.
\end{equation}
The convergence criterion for the shifted inverse power method is  
\begin{equation}\label{error_outer}
     \epsilon_\text{outer}=\left| \frac{\varpi^{(m+1)}}{\varpi^{(m)}}-1 \right|<10^{-5}.
\end{equation}

\subsection{CIS}

To find the VDF $\hat{f}^{(m+1)}$, the  kinetic equation  \eqref{Shakhov_sound_si_power} is solved by the conventional iteration scheme which treats the gain term in the previous inner iteration step $k$, while the rest terms are calculated in the current (inner) iteration step $k+1$:
\begin{equation}\label{Shakhov_sound_sip}
    \hat{f}^{(k+1)}=\frac{L^+_s\hat{f}^{(k)}-\hat{f}^{(m)}}{1+iKv_2+p},
\end{equation}
with the initial guess $\hat{f}=\hat{f}^{(m)}$. The inner iteration is considered converged when
$\epsilon_\text{inner}=\int \left| \hat{f}^{(k+1)}-\hat{f}^{(k)} \right|d\bm{v}<10^{-4}$. When the solution is converged, $\hat{f}^{(k+1)}$ is assigned to $\hat{f}^{(m+1)}$, which is then normalized, followed by another round of outer iteration described in Eq.~\eqref{Shakhov_sound_si_power}.

\begin{figure}[t!]
    \centering
    \includegraphics[trim={70 10 70 50},clip,width=0.48\linewidth]{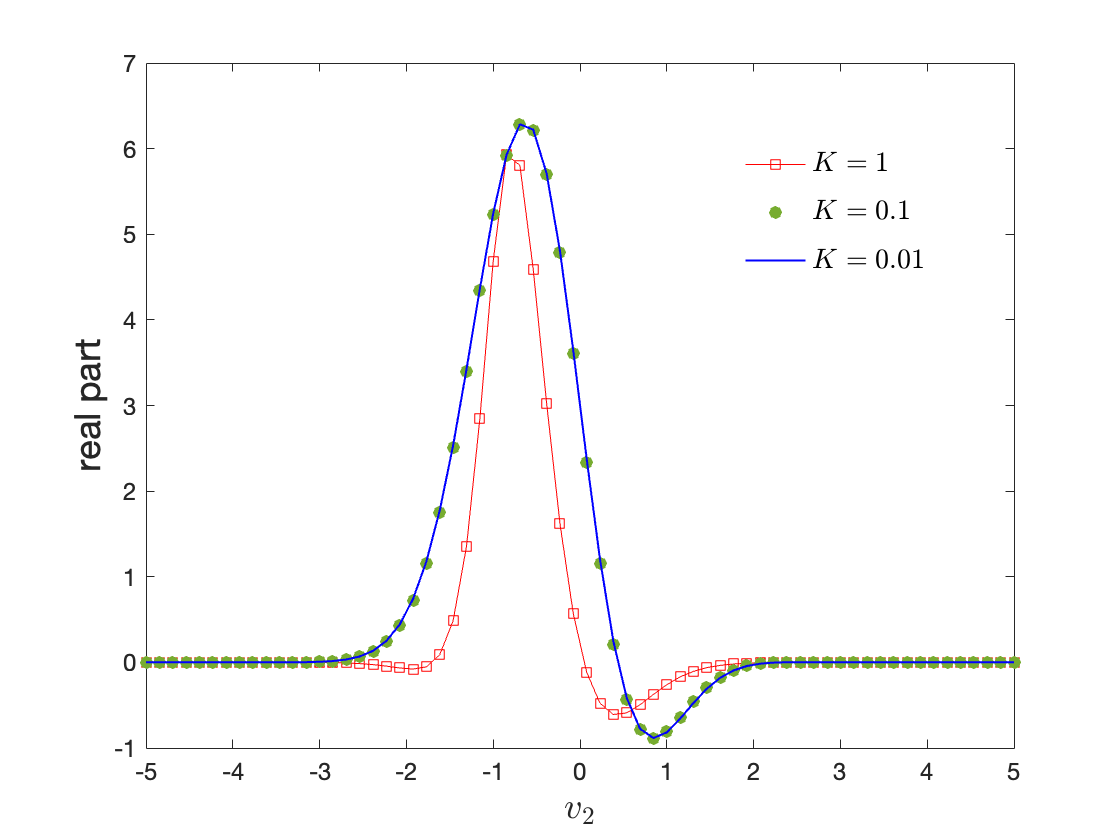}
    \includegraphics[trim={70 10 70 50},clip,width=0.48\linewidth]{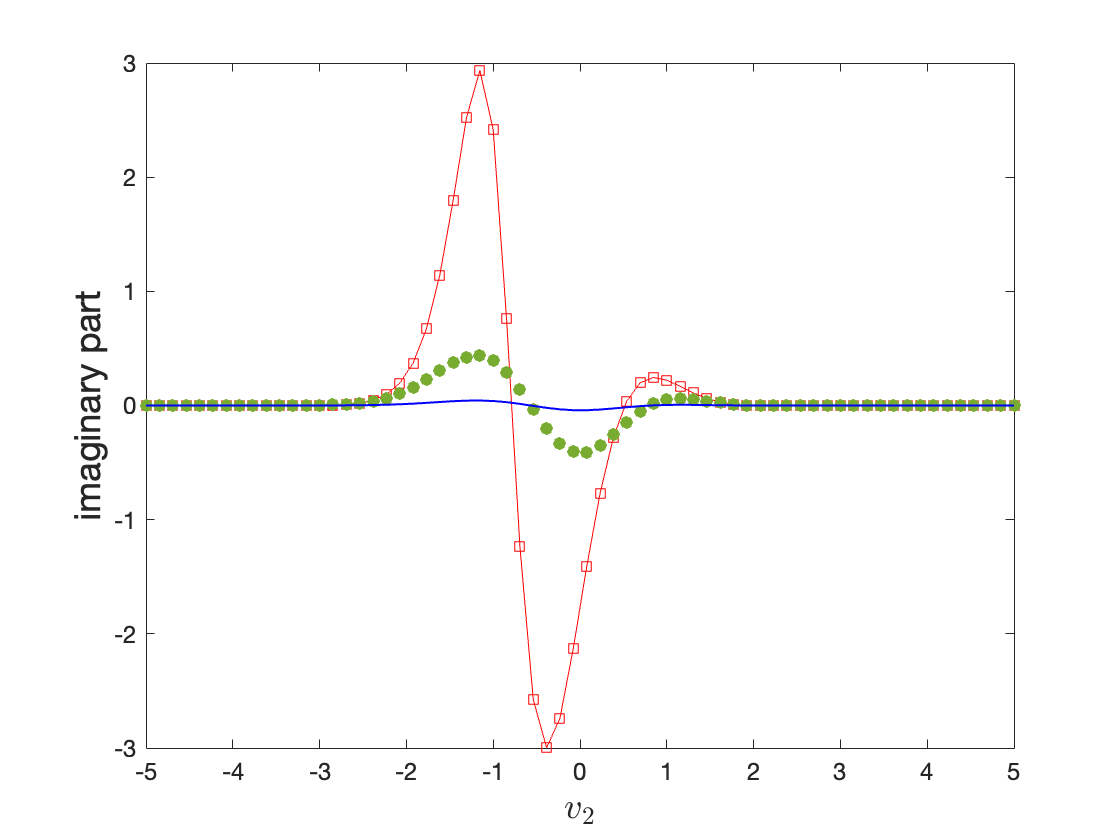}
    \caption{ Marginal VDFs $\int hdv_1dv_3$ in the planar sound wave obtained from the linearized Shakhov model. The shifted eigenvalue $p$ is chosen to match the imaginary part of the eigenvalue computed using the NS equations, that is, $p=9.13\times10^{-3}i, 9.11\times10^{-2}i$, and $0.68i$, when $K=0.01, 0.1$, and 1, respectively.
    }
    \label{fig:sound_vdf}
\end{figure}

\begin{figure}[t]
    \centering
    \includegraphics[trim={50 10 70 50},clip,width=0.6\linewidth]{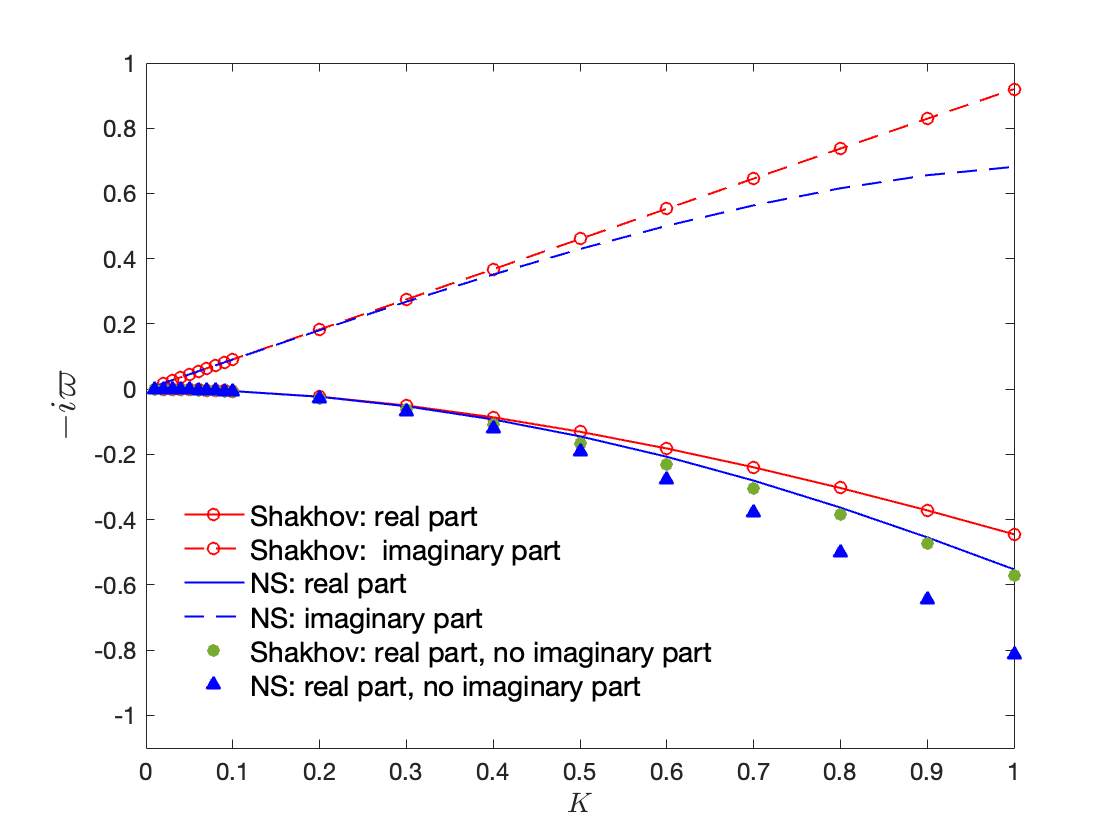}
    \caption{Eigenvalues in the planar sound wave as a function of the wavenumber. Lines are the traveling sound waves, while solid symbols are the standing sound wave.   }
    \label{fig:sound_eigvalue}
\end{figure}

Figure~\ref{fig:sound_vdf} shows eigenfunctions for $K=0.01,0.1$, and 1, corresponding to the continuum, slip, and transition flow regimes, respectively. This is because a smaller value of $K$ corresponds to a larger sound wavelength (or the actual characteristic flow length), resulting in a smaller effective Knudsen number.
It is observed that, when $K$ is small, the real part of the VDF remains nearly unchanged, while the imaginary part approximately scales with the value of $K$. As $K$ increases, the width of VDF gradually shrinks.

Figure~\ref{fig:sound_eigvalue} shows the eigenvalues $-i\varpi$ as a function of the wavenumber. These eigenvalues~\eqref{eig_power_itr} have been validated by the Rayleigh quotient
$-i\varpi={\int h(L-1-iKv_2)hdv}/{\int h^2dv}$. In the considered range of $K$, there are three sets of eigenvalues: one set is purely real, while the other two are complex conjugates, corresponding to standing and traveling sound waves, respectively. 
The eigenvalues obtained from the NS equations are also shown. It is observed that, when $K$ is small, the results from the Shakhov kinetic equation closely match those of the NS equations. However, as $K$ increases, rarefaction effects become significant, invalidating the constitutive relations of the NS equations and leading to discrepancies between the NS and Shakhov results.

The iteration numbers are summarized in Table~\ref{tab:sound_itr}. It is observed that the power method is efficient, converging within a few dozen outer iterations (denoted by $m$). However, the CIS method for solving the kinetic equation~\eqref{Shakhov_sound_sip} is considerably slower when the wavenumber is small; for example, $k \approx 200{,}000$ when $K = 0.01$. To accelerate convergence, the GSIS will be introduced, which can reduce the iteration count to $k = 10 \sim 30$.

\begin{table}[t!]
    \centering
    \caption{Comparison of iteration counts between the CIS and GSIS methods for the planar sound wave problem. In the calculation of traveling sound wave, the shifted eigenvalue $p$ is chosen to match the imaginary part of the eigenvalue computed using the NS equations, that is, $p=9.1285\times10^{-3}i, 9.1078\times10^{-2}i$, and $0.6816i$, when $K=0.01, 0.1$, and 1, respectively; in the calculation of standing sound wave, $p=0$. $\Sigma k$ is the total number of iterations, and the ratio indicates the speedup achieved by GSIS compared to CIS. Iteration steps of the GSIS-2 scheme in section~\ref{Further_acceleration} are also shown for comparison. }
    \begin{tabular}{ccccccc|cccccc}
    \hline
    & \multicolumn{6}{c|}{traveling sound} &  \multicolumn{6}{c}{standing sound}\\
    \hline
    & \multicolumn{2}{c}{CIS}  & \multicolumn{3}{c}{GSIS}  & GSIS-2 & \multicolumn{2}{c}{CIS}  & \multicolumn{3}{c}{GSIS} &GSIS-2\\
    \hline
        $K$    & $m$ & $\Sigma k$  & $m$ & $\Sigma k$ & ratio & $k$
        & $m$ & $\Sigma k$  & $m$ & $\Sigma k$ &  ratio &$k$\\     
        0.01 & 4 & 736534  & 4  & 89  & 8276 &7 &10 & 1459784 &9 & 341 & 4281 & 20\\    
        0.1  &  6 & 11505  & 6  & 98  & 117  & 11 & 8 & 11840 & 11 & 315 & 38  & 20 \\  
        1    & 21 & 767    & 21 & 250 & 3  &28 &18 & 554 & 18 & 231 & 2.4 & 29\\ \hline
    \end{tabular}
    \label{tab:sound_itr}
\end{table}

\subsection{GSIS}

In GSIS, the macroscopic equations are constructed to guide the quick evolution of the VDF~\cite{su2020can,su2020fast}. On multiplying Eq.~\eqref{Shakhov_sound_si_power} by 1, $v_2$, and $2v^2/3-1$ and integrating in the whole molecular velocity space, we obtain 
\begin{equation}\label{eq_sound_gsis}
\begin{aligned}
p\hat{\rho}+iK\hat{u}_2=0, \\
p\hat{u}_2+\frac{iK}{2}(\hat{\rho}+\hat{T}+\hat{\sigma}_{22})=0, \\
p\hat{T}+\frac{2iK}{3}(\hat{u}_2+\hat{q}_2)=0,
\end{aligned}
\end{equation}
where $\hat{\sigma}_{22}=2\int (v_2^2-c^2/3)\hat{f} d\bm{v}$.

In the continuum flow limit, according to the Chapman-Enskog expansion of the Boltzmann equation, the Newton law of stress and the the Fourier law of heat conduction are expressed as~\cite{chapman1990mathematical}
\begin{equation}
\begin{aligned}
    \hat{\sigma}^\text{NSF}_{22}=&-\frac{4}{3}\frac{\partial \hat{u}_{2}}{\partial {x_{2}}}, \quad
    \hat{q}^\text{NSF}_{2}=-\frac{5}{4\text{Pr}} \frac{\partial \hat{T}}{\partial x_2},
\end{aligned}
\end{equation}
so that the eigenvalues $-i\varpi$ is just the eigenvalues of the following matrix
\begin{equation}\label{NS_dispersion}
-\left[ \begin {array}{cccc} 0 & iK &0\\ \noalign{\medskip}
\frac{1}{2}iK & \frac{2}{3}K^2 & \frac{1}{2}iK
\\ \noalign{\medskip}0&\frac{2}{3}iK &
\frac{5}{6\text{Pr}}K^2
\end {array} \right].
\end{equation}

In rarefied gas flows, however, on top of the continuum constitutive relations, there are high-order constitutive relations, so that the stress and heat flux are expressed as
\begin{equation}\label{gsis_constitutive}
\begin{aligned}
\hat{\sigma}^{(k+1)}_{22} =& \hat{\sigma}^{(k+1),\text{NSF}}_{22}+\text{HoT}_{\sigma_{22}}, \\
\medskip
\hat{q}^{(k+1)}_2 =& \hat{q}^{(k+1),\text{NSF}}_2 +\text{HoT}_{q_2}, 
\end{aligned}
\end{equation}
where the high-order terms can only be obtained through numerical simulations, as no closed-form analytical expressions exist across the entire range of gas rarefaction, even after dozens of macroscopic equations are proposed in the history~\cite{LeiAiA}.

The GSIS is developed to boost the convergence when the effective Knudsen number is small, e.g., when $K$ is small in this sound wave problem.  
When $\hat{f}^{(k)}$ is known, rather than using the CIS~\eqref{Shakhov_sound_sip}, we first get the intermediate VDF in the following manner:
\begin{equation}\label{Shakhov_sound_GSIS}
    \hat{f}^{(k+1/2)}=\frac{L^+_s\hat{f}^{(k)}-\hat{f}^{(m)}}{1+iKv_2+p}.
\end{equation}
Then, the HoT terms are obtained as
\begin{equation}\label{gsis_constitutive_HoT}
\begin{aligned}
    \text{HoT}^{(k+1/2)}_{\sigma_{22}}=&2\int \hat{f}^{(k+1/2)} \left(v_2^2-\frac{c^2}{3}\right) d\bm{v} 
    -\hat{\sigma}^\text{NSF, k+1/2}_{22},\\
    \text{HoT}^{(k+1/2)}_{q_{2}}=&\int \hat{f}^{(k+1/2)} v_2\left({v^2}-\frac{5}{2}\right) d\bm{v}
    -\hat{q}^\text{NSF, k+1/2}_{2}.
\end{aligned}
\end{equation}
Second, substituting Eqs.~\eqref{gsis_constitutive} and~\eqref{gsis_constitutive_HoT} into Eq.~\eqref{eq_sound_gsis}, we can estimate the macroscopic quantities at the ($k$+1)-th inner iteration step as follows: 
\begin{equation}\label{GSIS_dispersion}
\left[ \begin {array}{cccc} 
p & iK &0 \\ \noalign{\medskip}
\frac{1}{2}iK & p+\frac{2}{3}K^2 & \frac{1}{2}iK
\\ \noalign{\medskip}
0 &\frac{2}{3}iK & p+\frac{5}{6\text{Pr}}K^2
\end {array} \right]
\left[ \begin {array}{cccc} 
\hat{\rho}^{(k+1)}  \\ \noalign{\medskip}
\hat{u}_2^{(k+1)} \\ \noalign{\medskip}
\hat{T}^{(k+1)}
\end {array} \right] =
-\left[ \begin {array}{cccc} \int \hat{f}^{(m)} d\bm{v} \\ \noalign{\medskip}
\int \hat{f}^{(m)}v_2 d\bm{v} +\frac{1}{2}iK\text{HoT}_{\sigma_{22}} \\ \noalign{\medskip}
\int \hat{f}^{(m)}\left(\frac{2}{3}v^2-1\right) d\bm{v}
+\frac{2}{3}iK\text{HoT}_{q_{2}}
\end {array} \right].
\end{equation}
Third, when Eq.~\eqref{GSIS_dispersion} is solved, the VDF is updated  to incorporate the change of the macroscopic quantities between the $(k+1)$-th step and the intermidiate step:
\begin{equation}\label{VDF_correction_sound}
\begin{aligned}
    \hat{f}^{(k+1)}=\hat{f}^{(k+1/2)}+& \left(\hat{\rho}^{(k+1)}-\hat{\rho}^{(k+1/2)}\right)f_{eq} \\
    +&\left(\hat{u}_2^{(k+1)}-\hat{u}_2^{(k+1/2)}\right) 2v_2f_{eq} \\
    +&\left(\hat{T}^{(k+1)}-\hat{T}^{(k+1/2)}\right)\left(v^2-\frac{3}{2}\right)f_{eq}.
\end{aligned}
\end{equation}
Finally, the whole process is repeated until convergence. 

We have verified that both CIS and GSIS yield the same eigenvalues and eigenfunctions. As shown in Table~\ref{tab:sound_itr}, GSIS significantly reduces the number of iterations, particularly in the continuum regime where $K$ is small. Hence the objective of restricting the inner iteration count to $k = 10 \sim 30$ is achieved.

It is worth mentioning that $1 + p$ appears in the denominator of Eq.~\eqref{Shakhov_sound_sip}. When $p$ becomes negative, the inner iteration may become divergent, say, when $K\gtrapprox1.5$ in the planar sound wave problem (see also the trend in Fig.~\ref{fig:sound_eigvalue}). As a result, the present method does not guarantee that all eigenvalues will be found. However, since our primary interest lies in the least stable mode and modes nearby, this issue is of little concern because we can avoid selecting values of $p$ with negative real parts.



\subsection{Linearized Boltzmann equation}

We now investigate the same problem using the linearized Boltzmann equation~\eqref{perturbation_Boltzmann}.  We consider monoatomic gases interacting through an inverse power-law potential and are particularly interested in whether this affects sound propagation. Under the base flow~\eqref{sound_base_flow}, the linearized Boltzmann equation becomes \cite{Wu2014JFM, Wu2015PoF}
\begin{equation}\label{Boltzmann_lin}
-i\varpi{}\hat{f}+iKv_2\hat{f}
=\mathcal{L}^+\hat{f}-\nu_{eq}(\bm{v})\hat{f}, 
\end{equation}
where the gain term  is
\begin{equation}\label{Boltzmann_lin_gain}
\mathcal{L}^+\hat{f}=\iint B(\theta,|\bm{v}-\bm{v}_\ast|)  [f_{eq}(\bm{v}')\hat{h}({\bm{v}}'_{\ast})+f_{eq}(\bm{v}'_\ast)\hat{h}({\bm{v}}')-f_{eq}(\bm{v})\hat{h}({\bm{v}}_\ast)]d\bm{\Omega} d{\bm{v}}_\ast,
\end{equation}
and the loss term is $\nu_{eq}(\bm{v})\hat{h}$, with 
$\nu_{eq}(\bm{v})=\iint{} B(\theta,|\bm{v}-\bm{v}_\ast|)f_{eq}(\bm{v}_{\ast}) d\Omega{d\bm{v}_\ast}$
being the equilibrium collision frequency.

\begin{table}[t]
\centering
\caption{Comparison of the eigenvalue $-i\varpi$ between the Boltzmann equation with various viscosity index $\omega$, the Shakhov models, and the Navier-Stokes equations, at $K=1$.}
\begin{tabular}{cccc}
\hline
Collision operator & viscosity index $\omega$ & Traveling wave & Standing wave \\
\hline
Boltzmann & 2 & $-0.3911 \pm 0.8745i$ & $-0.3874$ \\
Boltzmann & 1 & $-0.4526 \pm 0.9285i$ & $-0.4874$ \\
Boltzmann & 0.5 & $-0.4662 \pm 0.9718i$ & $-0.5543$ \\
Shakhov & - & $-0.4448 \pm 0.9200i$ & $-0.5702$ \\
Navier-Stokes & - & $-0.5523 \pm 0.6816i$ & $-0.8120$\\
\hline
\vspace{0.2cm}
\end{tabular}\label{Table_Boltzmann}
\end{table}

The inverse shifted power method and the GSIS remain unchanged, except that the linearized Boltzmann collision operator is solved with the fast spectral method~\cite{Wu2014JFM,Wu2015PoF} using 32, 128, and 32 uniform discrete velocity points in the $v_1, v_2$, and $v_3$  directions, respectively.
The Shakhov model, which is linearized with respect to global density and temperature, is independent of the intermolecular potential because its collision operator is overly simplified.
For the Boltzmann equation, Table~\ref{Table_Boltzmann} shows that, the eigenvalue of the traveling wave is very close to that of the Shakhov model, indicating that the potential has limited influence for most noble gases:
even for $\omega=2$ (close to a shielded Coulomb potential~\cite{Su2019JCP}), the relative difference is only approximately 7\%.
In the case of a standing sound wave, however, the decay rate is highly sensitive to the intermolecular potential: for $\omega=2$, the difference to the Shakhov model approaches 50\%. The different to that of the NS equation is even large,  reaching 110\%. 
These results indicate that the sound wave in the plasma will differ significantly from that in the noble gas.

\begin{figure}[t]
    \centering
    \includegraphics[trim={40 10 70 50},clip,width=0.48\linewidth]{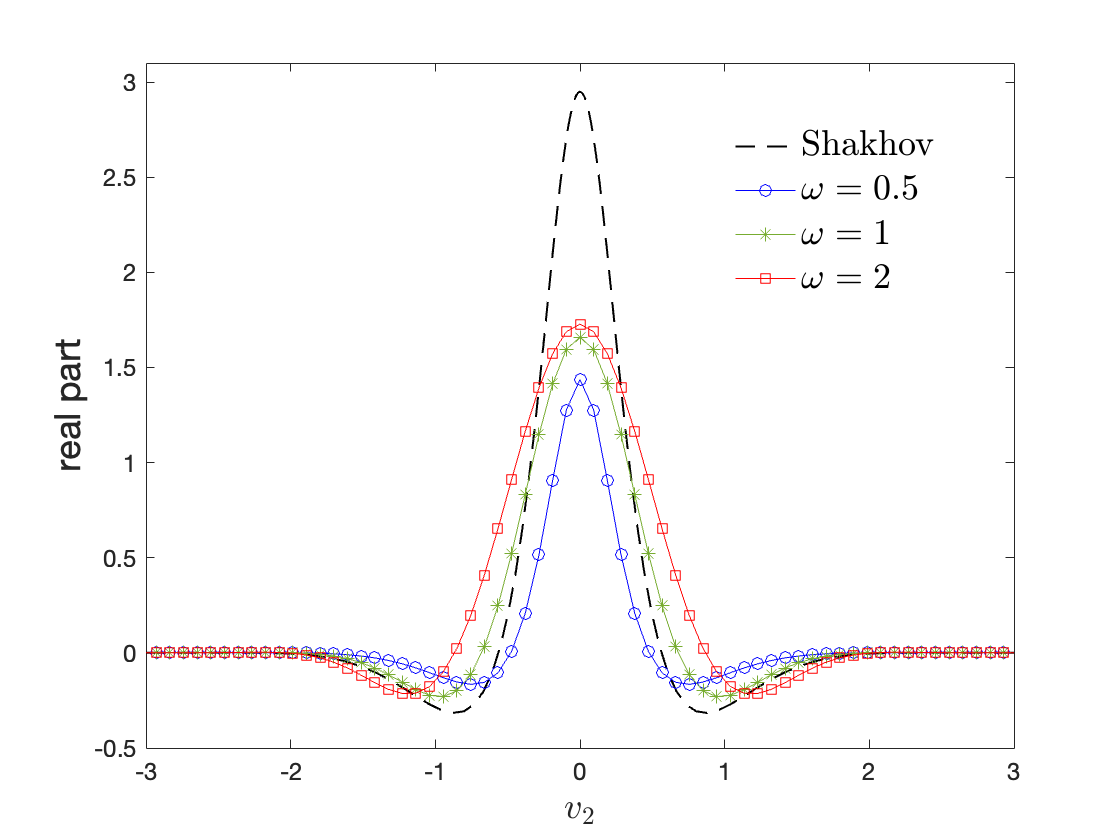}
    \includegraphics[trim={40 10 70 50},clip,width=0.48\linewidth]{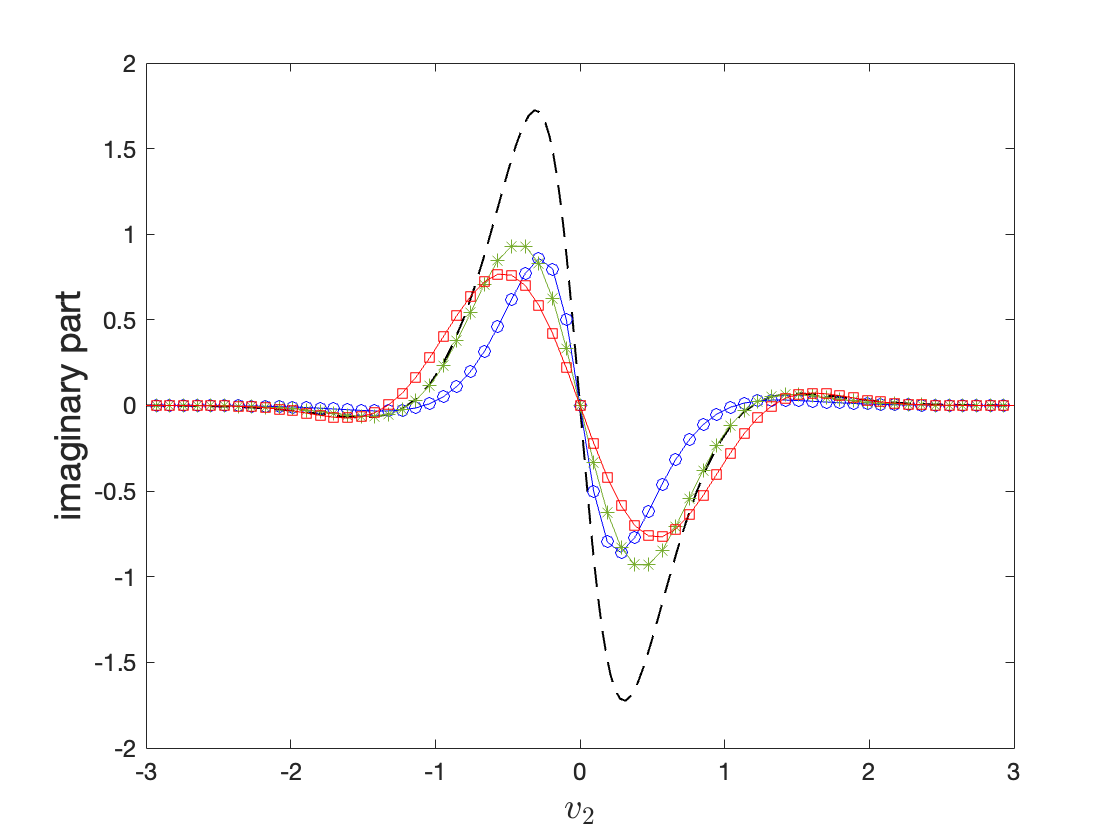}
    \caption{Marginal VDFs $\int hdv_1dv_3$ in the planar sound wave obtained from the linearized Boltzmann equation, when the shifted eigenvalue $p=0$. The wavenumber is $K=1$.
    }
    \label{fig:soundBoltzmann_vdf}
\end{figure}

Figure~\ref{fig:soundBoltzmann_vdf} presents the marginal VDFs obtained from the linearized Boltzmann equation for various values of the viscosity index. These correspond to the shielded Coulomb potential with an effective $\omega = 2$, the Maxwell gas with $\omega = 1$, and the hard-sphere gas with $\omega = 0.5$. As $\omega$ increases, the intermolecular potential decays more slowly as the intermolecular distance increases, and the width of the VDF broadens. A significant discrepancy from the Shakhov model is clearly observed.

\subsection{Further acceleration}\label{Further_acceleration}

The previous method consists of the shifted inverse power method in the outer iteration and GSIS in the inner iteration. In each inner iteration, $f^{(k+1)}$ is solved to convergence using Eq.~\eqref{Shakhov_sound_sip}, with $f^{(m)}$ held fixed. In the outer iteration, the converged $f^{(k+1)}$ is assigned to $f^{(m+1)}$ which is then scaled, followed by the next cycle of inner iterations. Therefore, it is possible to scale the value of $f^{(m)}$ within each inner iteration to reduce the total number of iterations. In this case, the outer and inner iterations are merged, and due to the instantaneous update of $f^{(m)}$, fewer total iterations are  required. We elaborate on the GSIS-2 scheme in  Algorithm~\ref{alg:GSIS-2}. Our numerical results in Table~\ref{tab:sound_itr} show that GSIS-2 reaches a converged solution within 30 iterations, making it approximately an order of magnitude faster than GSIS.

\begin{algorithm}[t]
    \caption{GSIS-2 based on the Shakhov model} 
    \label{alg:GSIS-2}
    \begin{algorithmic}[1]
        \Require{}wavenumber $K$; initial eigenvalue $\varpi^{(0)}=ip$; initial VDF $f^{(0)}=f^{(m)}=f_{eq}$;
        
        \Ensure{}converged eigenvalue and eigenfunction;

        \State Set $\epsilon_\text{outer}=1$;

        \State \textbf{While} $\epsilon_\text{outer}>10^{-5}$ \textbf{do}
        
        \State \hspace{1cm} get $f^{(k+1/2)}$ from Eq.~\eqref{Shakhov_sound_GSIS}; 

        \State \hspace{1cm} get $\hat{\rho}^{(k+1/2)}$,  $\hat{u}_2^{(k+1/2)}$, and $\hat{T}^{(k+1/2)}$ from Eq.~\eqref{MP};

         \State \hspace{1cm} solve Eq.~\eqref{gsis_constitutive_HoT} to get $\hat{\rho}^{(k+1)}$,  $\hat{u}_2^{(k+1)}$, and $\hat{T}^{(k+1)}$;
        
       \State \hspace{1cm} get $f^{(k+1)}$ from Eq.~\eqref{VDF_correction_sound}, then assign $f^{(k+1)}/\hat{\rho}^{(k+1)}$ to $f^{(m)}$;

       \State \hspace{1cm} the approximated eigenvalue is $\varpi^{(k+1)}=ip+i/{\hat{\rho}^{(k+1)}}$;

       \State \hspace{1cm} calculate $\epsilon_\text{outer}=\left| \frac{\varpi^{(k+1)}}{\varpi^{(k)}}-1 \right|$;
        
     \State \hspace{1cm} $k ++$;
          
    \State \textbf{end}
    \end{algorithmic}

Note that $\hat{\rho}^{(k+1)}$ in steps 6 and 7 can be replaced by $\hat{u}_2^{(k+1)}$ $\hat{T}^{(k+1)}$, or the VDF where the magnitude is largest.
\end{algorithm}


\section{Couette flow}\label{sec:Couette}

Consider a rarefied gas confined between two parallel plates located at $x_2=\pm1/2$, respectively. The plate at $x_2=1/2$ moving in the positive $x_1$ direction with speed $u_{w}$, while the plate at $x_2=-1/2$ moving in the negative $x_1$ direction with the same speed. Both plates are maintained at a normalized temperature of 1.


Assume that the $x_3$-direction is homogeneous and that the perturbation propagates along the $x_1$-direction with wavenumber $K$. Also assume that the wall velocity is much smaller than the most probable speed, so that the following base flow can be calculated by GSIS~\cite{Su2019JCP}:
\begin{equation}
    \rho_b(x_2)=T_b(x_2)=1, \quad 
    u_{b,1}=u_{b,1}(x_2), \quad 
    u_{b,2}(x_2)=0.
\end{equation}
Then, the kinetic LSE (the BGK mode with the Prandtl number being 1 is used for simplicity) can be greatly simplified to 
\begin{equation}\label{perturbation_Couette_BGK0}
\begin{aligned}
   -i\varpi \hat{f}+i{K}v_1\hat{f}+v_2\frac{\partial \hat{f}}{\partial x_2}
= &\delta_{rp} (L^+_p\hat{f}-\hat{f})
   +\delta_{rp} 
   \left(L^+_{s}f_b-f_b\right)\left[{\tilde{\rho}}+(1-\omega){\tilde{T}} \right], 
   \\ \noalign{\medskip}
   \text{with} \quad    
   L^+_p\hat{f}=& F_{eq}(1,{u}_{b,1},1)\left[
   \hat{\rho}+2(v_1-u_{b,1})\hat{u}_1
   +2v_2\hat{u}_2
   +\hat{T}\left(\frac{c_b^2}{T_b}-\frac{3}{2}\right)
   \right].
   \end{aligned}    
\end{equation}

To find the eigenvalue $-i\varpi$, the shifted inverse power method is applied, forming the following iterations:
\begin{equation}\label{BGK_Couette_si_power}
   \left(\delta_{rp} L^+_p-\delta_{rp}-p-i{K}v_1-v_2\frac{\partial \hat{f}}{\partial x_2}\right)\hat{f}^{(m+1)}
   +\delta_{rp} 
   \left(L^+_{s}f_b-f_b\right)\hat{\rho}^{(m+1)}=\hat{f}^{(m)},
\end{equation}
where $m$ is the outer iteration number and $p$ is the estimated eigenvalue.

To solve $\hat{f}^{(m+1)}$ given $\hat{f}^{(m)}$, we employ an inner iteration scheme, using the GSIS to accelerate convergence. Specifically, we first obtain the intermediate VDF  $\hat{f}^{(k+1/2)}$ by solving the following equation via the traditional sweeping algorithm:  
\begin{equation}\label{perturbation_Couette_BGK}
 \left( \delta_{rp}+p\hat{f}+i{K}v_1+v_2\frac{\partial }{\partial x_2} \right)  \hat{f}^{(k+1/2)}
= {\delta_{rp} L^+_p\hat{f}^{(k)}
   +\delta_{rp} 
   \left(L^+_{s}f_b-f_b\right)\hat{\rho}^{(k)} -\hat{f}^{(m)}},  
\end{equation}
where, in the numerical simulation, the spatial derivative is approximated using a second-order upwind finite difference scheme. That is to say, when $v_2>0$, the VDF is obtained from the VDF at the bottom wall at $x_2=-1/2$ to the top wall at $x_2=1/2$. When $v_2<0$, the VDF is obtained from the VDF at the top wall to the bottom wall. 

The core of GSIS is to formulate a macroscopic synthetic equation that guides the evolution of the VDF. According to Eq.~\eqref{perturbation_quantities}, we respectively multiply Eq.~\eqref{BGK_Couette_si_power} with $1$, $2(v_1-u_{b,1})$, $2v_2$, and $c_b^2-3/2$, obtaining 
\begin{equation}\label{synthetic_Couette}
    \begin{aligned}
p\hat{\rho}+i{K}\left(\hat{u}_1+u_{b,1}\hat{\rho}\right)+\frac{\partial \hat{u}_2}{\partial x_2}  =&-\int \hat{f}^{(m)} d\bm{v},\\  \medskip
2p\hat{u}_1+i{K}\left(\hat{\sigma}_{11}+\hat{\rho} +\hat{T} +2u_{b,1}\hat{u}_1\right)
       + \frac{\partial\hat{\sigma}_{12}}{\partial x_2}
       +2\frac{\partial{}u_{b,1}}{\partial x_2}\hat{u}_2=&-2\int (v_1-u_{b,1}) \hat{f}^{(m)} d\bm{v},\\ 
        \medskip
2p\hat{u}_2+  i{K}\left(\hat{\sigma}_{12}+2u_{b,1}\hat{u}_2\right)
        +\frac{\partial}{\partial x_2}(\hat{\sigma}_{22}+\hat{\rho} +\hat{T})=&-2\int v_2\hat{f}^{(m)} d\bm{v}, \\
        \medskip
 \frac{3}{2}p\hat{T}+iK\left(\hat{q}_1+\hat{u}_1+\frac{3}{2}u_{b,1}\hat{T}\right)  
 +\frac{\partial}{\partial{x_2}}(\hat{q}_2+\hat{u}_2) +\frac{\partial{u_{b,1}}}{\partial{}x_2}\hat{\sigma}_{12}=& -\int \left(c_b^2-\frac{3}{2}\right)\hat{f}^{(m)} d\bm{v}.   
    \end{aligned}
\end{equation}
For the inner iteration~\eqref{perturbation_Couette_BGK}, the Newton law of stress and high-order constitutive relations are used
\begin{equation}\label{sigma_first_HoT}
\begin{aligned}
    \hat{\sigma}^{(k+1)}_{ij}=\hat{\sigma}^{(k+1),\text{NSF}}_{ij} +\underbrace{ \hat{\sigma}^{(k+1/2)}_{ij}-  \hat{\sigma}^{(k+1/2),\text{NSF}}_{ij}}_{\text{HoT}^{k+1/2}_{\sigma_{ij}}}, \\
   \hat{q}^{(k+1)}_{i}=\hat{q}^{(k+1),\text{NSF}}_{i} +\underbrace{ \hat{q}^{(k+1/2)}_{i}-  \hat{q}^{(k+1/2),\text{NSF}}_{i}}_{\text{HoT}^{k+1/2}_{q_{i}}},   
\end{aligned}
\end{equation}
where
$\hat{\sigma}^{(k+1),\text{NSF}}_{ij}=-{\delta^{-1}_{rp}}\left[\frac{\partial \hat{u}_i}{\partial x_j}+\frac{\partial \hat{u}_j}{\partial x_i}-\frac{2}{3}\frac{\partial \hat{u}_{k'}}{\partial x_{k'}} \delta_{ij}\right]$, and the Einstein summation is used for index $k'=1,2$ or 3.
In order to avoid ambiguity, we give the detailed expressions here:
\begin{equation}
    \begin{aligned}
           \hat{\sigma}^{(k+1),\text{NSF}}_{11}=&-\frac{1}{\delta_{rp}}\left[2\frac{\partial \hat{u}_1}{\partial x_1}
           -\frac{2}{3}\left( \frac{\partial \hat{u}_1}{\partial x_1}+\frac{\partial \hat{u}_2}{\partial x_2}\right)\right]
           =-\frac{1}{\delta_{rp}}\left[\frac{4}{3}i{K}\hat{u}_1
           -\frac{2}{3}\frac{\partial \hat{u}_2}{\partial x_2}\right],\\ \medskip
    \hat{\sigma}^{(k+1),\text{NSF}}_{22}=&-\frac{1}{\delta_{rp}}\left[2\frac{\partial \hat{u}_2}{\partial x_2}
           -\frac{2}{3}\left( \frac{\partial \hat{u}_1}{\partial x_1}+\frac{\partial \hat{u}_2}{\partial x_2}\right)\right]
           =-\frac{1}{\delta_{rp}}\left[\frac{4}{3}\frac{\partial \hat{u}_2}{\partial x_2}-\frac{2}{3}i{K}\hat{u}_1 \right],\\ \medskip
\hat{\sigma}^{(k+1),\text{NSF}}_{12}=&-\frac{1}{\delta_{rp}}
\left[\frac{\partial \hat{u}_1}{\partial x_2}+\frac{\partial \hat{u}_2}{\partial x_1} \right]
=-\frac{1}{\delta_{rp}}
\left[\frac{\partial \hat{u}_1}{\partial x_2}+i{K}\hat{u}_2 \right],
    \end{aligned}
\end{equation}
where, as Eq.~\eqref{plane_wave}, $\partial/\partial x_1$ is replaced by $i{K}$. Similarly, 
\begin{equation}
    \hat{q}^{(k+1),\text{NSF}}_1=-\frac{5}{4}iK\hat{T}, \quad 
    \hat{q}^{(k+1),\text{NSF}}_2=-\frac{5}{4}\frac{\partial\hat{T}}{\partial x_2}.
\end{equation}

In the numerical simulation, the perturbation density $\hat{\rho}$ is obtained by solving the first equation in Eq.~\eqref{synthetic_Couette}:
\begin{equation}\label{rho_equation}
\hat{\rho}=-\frac{1}{p+iKu_{b,1}}
\left(\int \hat{f}^{(m)} d\bm{v}+iK\hat{u}_1 +
\frac{\partial \hat{u}_2}{\partial x_2}\right),
\end{equation}
which is then substituted into the second and third equations of Eq.~\eqref{synthetic_Couette}. Therefore, we have three linear ordinary differential equations for $\hat{u}_1$, $\hat{u}_2$, and $\hat{T}$.  A five-point central finite-difference scheme is employed at interior grid points, while near the top (bottom) boundary, a five-point backward (forward) finite-difference scheme is used. The velocity and temperature at the boundaries are determined from the kinetic equation~\eqref{perturbation_Couette_BGK}. 

When the velocity and temperature in Eq.~\eqref{synthetic_Couette} are solved at the $(k+1)$-th inner iteration, the density is obtained from  Eq.~\eqref{rho_equation}. Then the VDF are updated as follows: 
\begin{equation}\label{Couette_correction}
\begin{aligned}
    \hat{f}^{(k+1)}=\hat{f}^{(k+1/2)}+& {0.2}\underline{\left(\hat{\rho}^{(k+1)}-\hat{\rho}^{(k+1/2)}\right)f_{eq}}
    +\left(\hat{T}^{(k+1)}-\hat{T}^{(k+1/2)}\right)\left(c_b^2-\frac{3}{2}\right)f_{eq}
   \\
   \medskip
    +&2\left(\hat{u}_1^{(k+1)}-\hat{u}_1^{(k+1/2)}\right) (v_1-u_{b,1})f_{eq} 
    +2\left(\hat{u}_2^{(k+1)}-\hat{u}_2^{(k+1/2)}\right) v_2f_{eq}.
\end{aligned}
\end{equation}
Note that, under normal circumstances, the coefficient in front of the underlined term should be 1. However, in the finite difference method, $\hat{\rho}$ cannot be accurately obtained from Eq.~\eqref{rho_equation} near the boundary. Therefore, an empirical damping factor of 0.2 is introduced to ensure numerical stability. Once the solution converges and $\hat{\rho}^{(k+1)} = \hat{\rho}^{(k+1/2)}$, this damping does not affect the final accuracy.

Finally, although we describe the method as the shifted inverse power method in the outer iteration and GSIS in the inner iteration, the GSIS-2 algorithm is used in practice—similar to the approach presented in section~\ref{Further_acceleration}. Specifically, after obtaining $\hat{f}^{(k+1)}$ from Eq.~\eqref{Couette_correction}, the new perturbed density $\hat{\rho}$ is computed. Then, the VDF in the power method is normalized as
\begin{equation}
    \hat{f}^{(m)}(x_2,\bm{v})=\frac{\hat{f}^{(k+1)}(x_2,\bm{v})}{\int \hat{\rho}(x_2) dx_2},
\end{equation}
while the eigenvalue is calculated as 
\begin{equation}
    -i\varpi=p+\frac{1}{\int \hat{\rho}(x_2) dx_2}.
\end{equation}

\subsection{The transition regime}

\begin{figure}[t]
    \centering
    \includegraphics[trim={40 30 70 70}, width=1\linewidth]{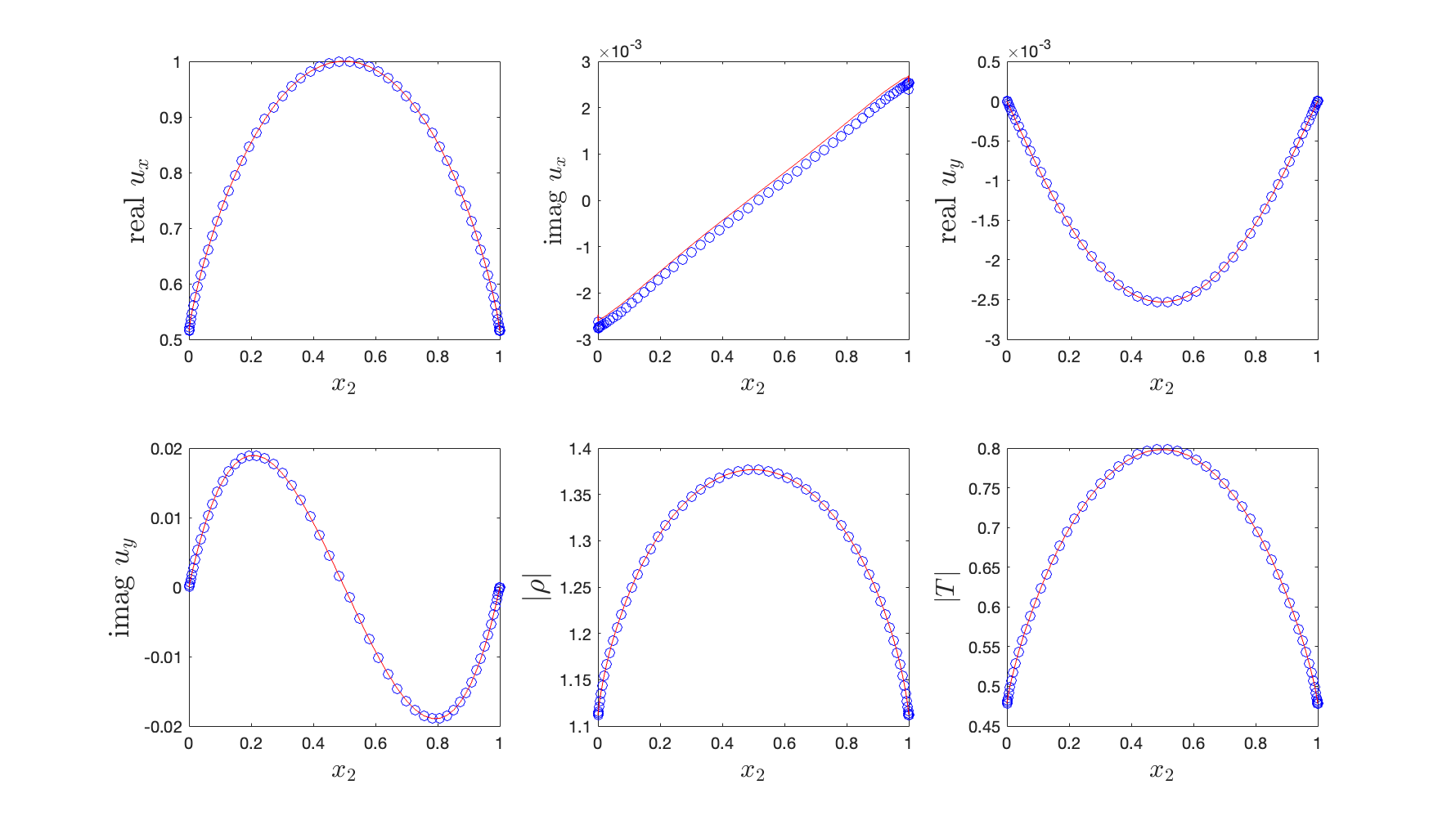}
    \caption{Profiles of the eigenfunctions of the velocity, density, and temperature, when $\delta_{rp}=1$ and $K=1$. Results with lines and circles are obtained from the CIS and GSIS, respectively. }
    \label{fig:Couettedelta1K1}
\end{figure}

We consider Couette flow of Maxwell gas ($\omega=1$) with a rarefaction parameter $\delta_{rp} = 1$ and a wall velocity equal to 0.01 times the most probable speed. The molecular velocity component $v_1$ is discretized using a 24-point Gauss–Hermite quadrature, $v_3$ is discretized using a 4-point Gauss–Hermite quadrature, and $v_2$ is discretized using $N_v = 32$ non-uniform discrete points as~\cite{Wu2014JFM} 
    \begin{equation}\label{nonuniform_v}
v=\frac{5}{(N_v-1)^3}(-N_v+1,-N_v+3,\cdots,{N_v-1})^3,
\end{equation}
in order to capture the discontinuities in the VDF around $v_2\sim0$. The spatial variable $x_2$ is also discretized by $N_s=60$ non-uniform points as 
\begin{equation}\label{spatial_d}
x_2=(10-15s+6s^2)s^3, 
\quad
s=(0,1,\cdots,N_s)/N_s,
\end{equation}
in order to capture the sharp variation of the VDF near the solid wall. 

Starting from the initial VDF $\hat{f} = f_{eq}$ and an estimated eigenvalue of $p = 0.5i$, we run both the CIS and GSIS-2 methods until the relative error in the eigenvalue between two consecutive iterations falls below $10^{-5}$, when the perturbation wavenumber is $K=1$. The required numbers of iterations are 49 for CIS and 45 for GSIS, respectively. This small difference arises because CIS is already efficient in this case, as binary collisions are relatively infrequent. The resulting eigenvalue is $-i\varpi = -0.6564 + 0.0002i$. Since, at large Knudsen numbers, the eigenvalue is known to be a real negative number~\cite{Bi2025CNSNS}, this suggests that our method achieves an accuracy on the order of $10^{-4}$. It is worth noting that, since the shifted inverse power method finds the eigenvalue closest to the shift value $p$, the resulting solution corresponds to the mode with the minimum damping rate—sometimes referred to as the least stable mode.

Figure~\ref{fig:Couettedelta1K1} shows the corresponding eigenfunctions. These functions are normalized such that the phase of $\hat{u}_x$ is set to zero at the point where $|\hat{u}_x|$ reaches its maximum. It can be observed that, due to significant rarefaction effects in the transition regime, the horizontal velocity exhibits a pronounced velocity slip. The imaginary part of $\hat{u}_x$ and real part of $\hat{u}_y$ is approximately three orders of magnitude smaller than its real part; this is due to the numerical error caused by the velocity discretization and convergence criterion, as they should be zero. Similarly, the real part of vertical velocity $\hat{u}_y$ remains very small throughout. 
Our numerical results suggest that the perturbed temperature is not zero, which differs from the assumption in Ref.~\cite{Zou2023JFM}. Furthermore, although the absolute density and temperature both exhibit parabolic profiles, their variations are out of phase, resulting in $\hat{\rho} + \hat{T}$ remaining nearly constant across the domain.

\subsection{The slip regime}

\begin{figure}[t]
    \centering
    \includegraphics[trim={40 30 70 70}, width=0.78\linewidth]{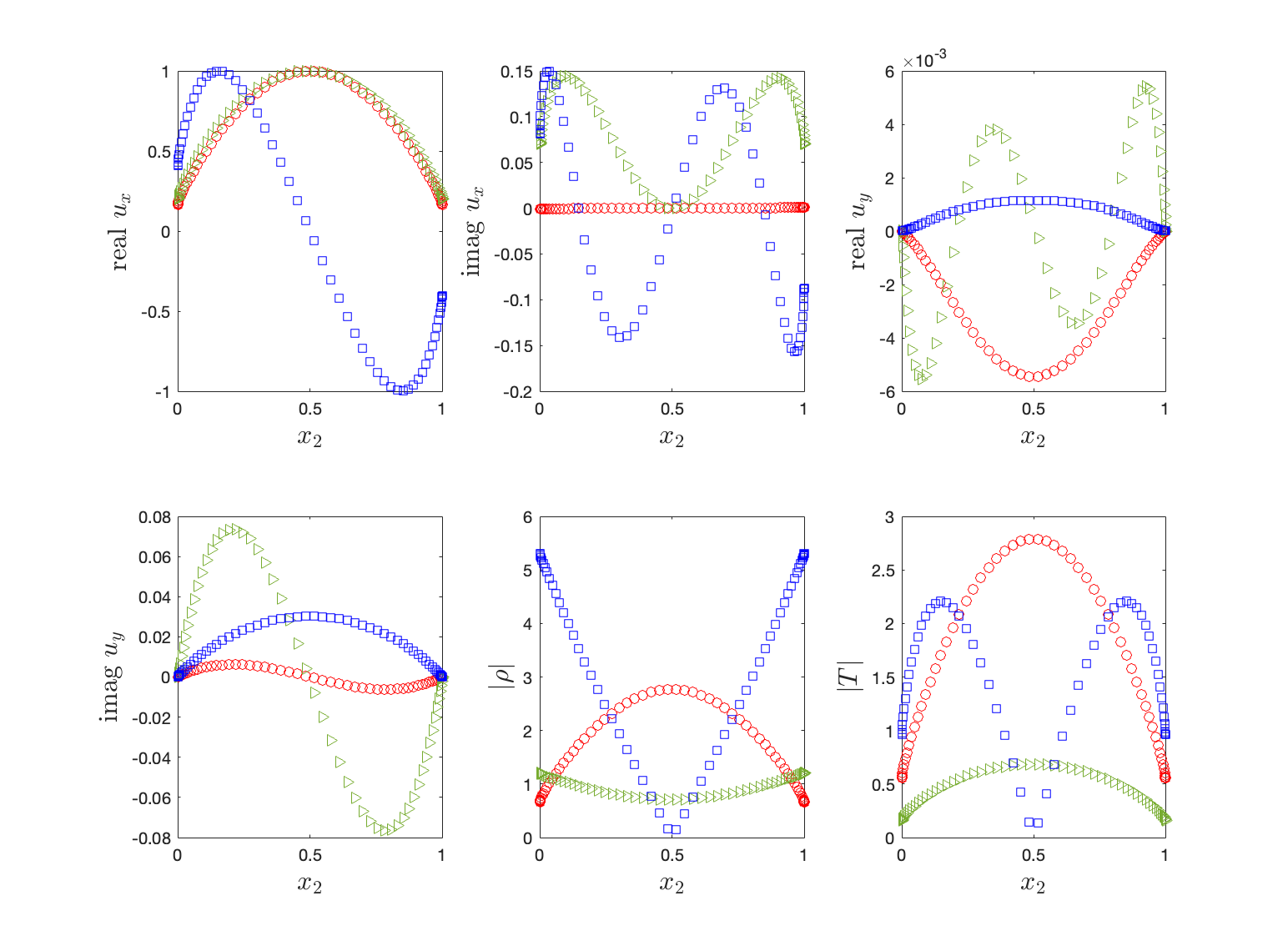}
    \includegraphics[width=0.19\linewidth]{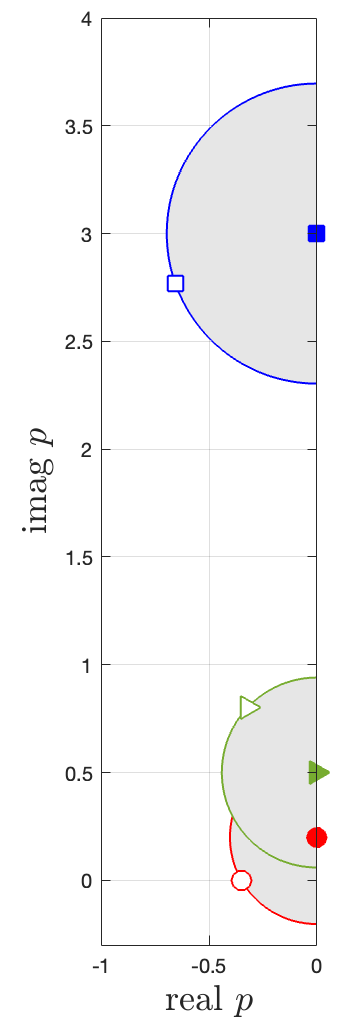}
    \caption{Profiles of the eigenfunctions of velocity, density, and temperature for $\delta_{rp} = 10$ and $K = 1$. Circles, triangles, and squares represent the results obtained using GSIS with initial guesses of $p = 0.2i$, $0.5i$, and $3i$, respectively. The open symbols in the right column indicate the corresponding true eigenvalues: $-i\varpi=-0.3497$, $-0.3208+0.8017i$, and $-0.6566+2.7695i$.
    The gray region denotes the absence of any eigenvalues other than the three identified. Note that the real and imaginary parts of $\hat{u}_y$ have been scaled down by factors of 1000 and 100, respectively, to facilitate clearer comparison.
    }
    \label{fig:Couettedelta10K1}
\end{figure}

We then consider the same Couette flow with a rarefaction parameter of $\delta_{rp} = 10$. In Fig.~\ref{fig:Couettedelta10K1}, three eigenvalues and their corresponding eigenfunctions are identified using appropriate initial guesses for the eigenvalues. When $p = 0.2i$, we obtain a purely damped, non-oscillatory mode with the eigenvalue $\i\varpi = -0.3497$. Regarding the perturbation velocity, similar to the case with $\delta_{rp} = 1$, only the real part of $\hat{u}_x$ is dominant. 

The properties of the inverse power method indicate that there are no additional eigenvalues within the red circle shown in the right column of Fig.~\ref{fig:Couettedelta10K1}. Therefore, we select a second initial guess of $p = 0.5i$. After 170 iterations and 6.5 seconds on a laptop using MATLAB, we identify another eigenvalue, $-i\varpi = -0.3208 + 0.8017i$. In contrast, the CIS needs 2924 iterations and 88.5 seconds to produce the same solution. This eigenvalue clearly corresponds to a damped oscillatory mode. The real part of $\hat{u}_x$ nearly overlaps with that of the purely damping mode, while its imaginary part exhibits a double-peak structure, with a magnitude 10 times smaller. Meanwhile, the profile of $|\hat{\rho}|$ shows a reversed trend compared to that of the purely damping mode. This mode is, in fact, the least stable one, as it has the smallest magnitude of the real part of the eigenvalue.

By further increasing the initial guess of $p$, we identify a third eigenvalue: $-i\varpi = -0.6566 + 2.7695i$. The iteration is efficient, requiring only 110 steps and taking just 4.5 seconds on a laptop using MATLAB.  This mode decays rapidly but has a high oscillation frequency. The magnitude of $\hat{u}_y$ is of the same order as $\hat{u}_x$, both being approximately three orders of magnitude larger than those of the previous two modes. The perturbed density and temperature change sign at the midpoint between the two plates. 

The conjugate modes can be obtained by replacing $p$ with $-p$.

\subsection{The continuum regime}

\begin{figure}[t]
    \centering
    \includegraphics[width=0.9\linewidth]{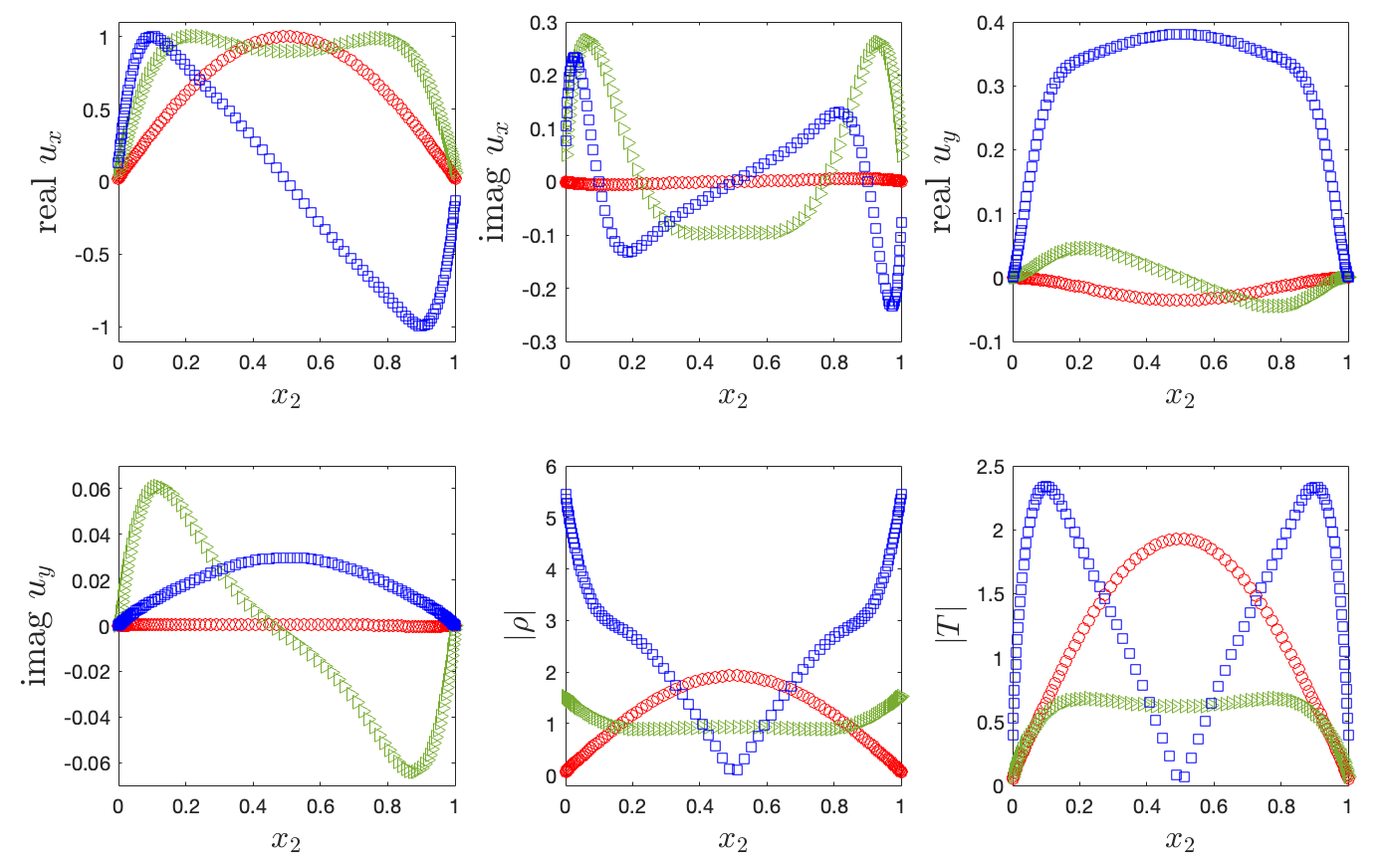}
    \caption{Profiles of the eigenfunctions for velocity, density, and temperature at $\delta_{rp} = 100$ and $K = 1$. Circles, triangles, and squares represent the results obtained using GSIS-2 with initial guesses of $p = 0.1$, $i$, and $3i$, respectively. The corresponding true eigenvalues are $-i\varpi = -0.0517$, $-0.0875 + 0.8487i$, and $-0.1705 + 2.9166i$. To facilitate clearer comparison, the density and temperature profiles (circles) have been scaled down by a factor of 10, and the imaginary part of $\hat{u}_y$ (squares) has also been scaled down by a factor of 10.}
    \label{fig:Couettedelta100K1}
\end{figure}

Finally, we consider the Couette flow at a rarefaction parameter of $\delta_{rp} = 100$. Since the VDF is relative smooth in the $v_1$ direction, the molecular velocity space is also discretized by 4-point Gauss-Hermite quadrate.

Figure~\ref{fig:Couettedelta100K1} presents the initial guess for the eigenvalues, along with the three eigenvalues and corresponding eigenfunctions obtained using GSIS-2. The CIS is too slow to reach a converged solution. Interestingly, unlike in the slip and transition flow regimes, the least stable mode in this case is a purely damped mode with $-i\varpi = -0.0517$. For the damped oscillatory mode with the second largest growth rate, i.e., $-i\varpi = -0.0875 + 0.8487i$, we observe a marked difference in the eigenfunctions compared to those in the slip flow regime. Specifically, the real part of $\hat{u}_x$ exhibits two distinct peaks, while the imaginary part of $\hat{u}_x$ shows a nearly flat profile in the central region of the computational domain. For the other eigenfunctions, when compared with those in the slip flow, the profiles appear less smooth—likely due to the stronger rarefaction effects in the slip regime, which tend to smooth out such variations.
\section{Conclusions and outlooks}\label{sec:conclusion}

In summary, we have proposed a novel linear stability analysis for the Boltzmann equation and simplified kinetic model equations. This method combines the shifted inverse power technique for solving eigenvalue problems with a general synthetic iterative scheme that ensures fast convergence and asymptotic-preserving properties for steady-state solutions. As a proof of concept, the accuracy and efficiency of the proposed approach have been demonstrated through its application to the planar sound wave and Couette flow. Although our method captures only the least stable mode and a few nearby modes—unlike the work of Zou \textit{et al} who computed all eigenvalues and eigenfunctions by reformulating the eigenvalue problem as a large linear system~\cite{Zou2023JFM}—it offers superior computational efficiency and is well-suited for problems where only the dominant modes are of interest. 

Specifically, while directly solving the eigenvalue problems for three-dimensional rarefied gas flows—where the velocity distribution function is defined in a seven-dimensional phase space—is computationally prohibitive, our method will remain applicable. To provide a rough estimate: the number of iterations required to solve the kinetic equation using our approach is on the order of a few hundred, which is only slightly higher than that needed for solving 3D rarefied gas flows using the parallel solver developed by Zhang \textit{et al}~\cite{zhang2024efficient}. Based on this, we estimate that the least stable mode and a few nearby modes for hypersonic flows around the Apollo reentry capsule or X-38-like space vehicles can be obtained within a few hours using 512 CPU cores, when about 1 million spatial cells and 10,000 discretized velocities are used. In other words, our method enables feasible linear stability analysis of near-space hypersonic flows based on gas kinetic models, where the validity of the Navier–Stokes equations may break down in the multiscale flows.

Furthermore, although we have only considered a monatomic gas, the present method can be readily extended to kinetic systems with internal degrees of freedom to study the stability of non-equilibrium gas flows—such as carbon dioxide flows, where a high bulk viscosity significantly influences flow transitions~\cite{Li2016AIAA}. This may have applications in Mars landing missions.

Finally, we would like to highlight some limitations of the present method. Since it relies on the power method to compute the eigenpairs, convergence is only linear and can be slow—particularly when the dominant eigenvalue is not well separated from the others. One possible solution is to combine the power method with the Rayleigh quotient iteration to accelerate convergence. Another possible approach draws inspiration from techniques used in neutron transport~\cite{Alcouffe01101977,ADAMS20023}, where high-order constitutive relations are incorporated into effective transport coefficients. The eigenpairs can then be obtained by solving the LSE of the resulting Navier–Stokes equations with effective shear viscosity and thermal conductivity.
However, in rarefied gas flows, many eigenvalues must be found, and the selected eigenvalue may shift during iteration as the effective constitutive relations change. Therefore, several challenges remain to be addressed, which will be explored in future work. Nonetheless, we emphasize that the present study represents a significant step forward in solving the linear stability analysis of the Boltzmann kinetic equation for general 3D flows.

\section*{Acknowledgments} 
This work is supported by the National Natural Science Foundation of China (12450002). The author thanks Dr. Xiafeng Zhou for pointing out during a conference that the conventional iterative scheme shares core steps with the inverse power method for $k$-eigenvalue problems in neutron transport; this work builds upon this previously overlooked insight by the author.


\bibliographystyle{elsarticle-num}

\bibliography{ref}

\end{document}